\documentclass[useAMS,usenatbib]{mn2e}
\usepackage{amsmath}
\usepackage{amssymb}
\usepackage{graphicx}
\usepackage{epsfig}
\makeatletter{}
\def\Mpch{\mbox{$\,h^{-1}$Mpc}}
\def\kMpch{\mbox{$h$Mpc$^{-1}$}}
\def\Mpc{\mbox{Mpc}}
\def\kbox{\mbox{$k_{\rm box}$}}

\def\Gpch{\mbox{$\,h^{-1}$Gpc}}

\def\M200{\mbox{$M_{\rm 200 }$}}
\def\Msunh{\mbox{$h^{-1}M_\odot$ }}

\def\R200{\mbox{$R_{\rm 200 }$}}

\def\V200{\mbox{$V_{\rm 200 }$}}

\def\GLAM{\textsc{glam}\,\,}

\newcommand{\Ng}{\mbox{$N_{\rm g}$}}

\newcommand{\lsim}{\mbox{${\,\hbox{\hbox{$ < $}\kern -0.8em \lower 1.0ex\hbox{$\sim$}}\,}$}}
\newcommand{\gsim}{\mbox{${\,\hbox{\hbox{$ > $}\kern -0.8em \lower 1.0ex\hbox{$\sim$}}\,}$}}

\def\beqn{\vspace{2mm}
\begin{eqnarray}} 
\def\eeqn{\vspaceg{2mm} 
\end{eqnarray}}

\newcommand{\be}{\begin{equation}}
\newcommand{\ee}{\end{equation}}
\newcommand{\ba}{\begin{eqnarray}}
\newcommand{\ea}{\end{eqnarray}}
\newcommand{\brr}{\begin{array}}
 
\newcommand{\err}{\end{array}}
\newcommand{\bc}{\begin{center}}
\newcommand{\ec}{\end{center}}

\title[Effects of long-waves in large surveys]
{Effects of long-wavelength fluctuations in large galaxy surveys}

\makeatletter{}\author[Klypin \& Prada]
  {Anatoly~Klypin$^{1,2}$ and Francisco~Prada$^{3}$
   \vspace{0.2cm}\\ 
  $^1$ Astronomy Department, New Mexico State University, Las Cruces, NM, USA\\
  $^2$ Department of Astronomy, University of Virginia, Charlottesville, VA, USA\\
  $^3$ Instituto de Astrof\'{\i}sica de Andaluc\'{\i}a (CSIC), Glorieta de 
     la Astronom\'{\i}a, E-18080 Granada, Spain \\
}

\author[Klypin \& Prada]{Anatoly~Klypin$^{1,2}$\thanks{E-mail: aklypin@nmsu.edu} and Francisco~Prada$^{3}$\\
 \vspace{-0.2cm}\\
$^1$ Astronomy Department, New Mexico State University, Las Cruces, NM, USA\\
$^2$ Department of Astronomy, University of Virginia, Charlottesville, VA, USA\\
$^3$ Instituto de Astrof\'{\i}sica de Andaluc\'{\i}a (CSIC), Glorieta de 
     la Astronom\'{\i}a, E-18080 Granada, Spain \\
}

\begin{document}
\maketitle
\label{firstpage}
\begin{abstract}
  In order to capture as much information as possible large galaxy
  surveys have been increasing their volume and redshift depth. To
  face this challenge theory has responded by making cosmological
  simulations of huge computational volumes with equally increasing
  the number of dark matter particles and supercomputing
  resources. Thus, it is taken for granted that the ideal situation is
  when a single computational box encompasses the whole 
  volume of the observational survey, e.g.,
  $\sim 50\, h^{-3}{\rm Gpc}^3$ for the DESI and Euclid surveys. Here we
  study the effects of missing long-waves in a finite volume using
  several relevant statistics: the abundance of dark matter halos, the
  PDF, the correlation function and power spectrum, and covariance
  matrices. Finite volume effects can substantially modify the results
  if the computational volumes are less than $\sim (500\Mpch)^3$.
  However, the effects become extremely small and practically can be
  ignored when the box-size exceeds $\sim 1$\,Gpc$^3$. We find that
  the average power spectra of dark matter fluctuations show
  remarkable lack of dependence on the computational box-size with
  less than 0.1\% differences between $1\Gpch$ and $4\Gpch$ boxes. No
  measurable differences are expected for the halo mass functions for
  these volumes. The covariance matrices are scaled trivially with
  volume, and small corrections due to super-sample modes can be
  added. We conclude that there is no need to make those extremely
  large simulations when a box-size of $1-1.5\Gpch$ is sufficient to
  fulfil most of the survey science requirements.
\end{abstract}


\makeatletter{}\section{Introduction}
\label{sec:intro}

Large-scale galaxy surveys such as the existing 2dFGRS \citep{Hawkins2003}, the SDSS 
(\citep[e.g.,][]{Anderson2012,eBOSS}, and the upcoming DESI \citep{DESI2016}, 
Euclid \citep{Euclid}, LSST \citep{LSST}, and WFIRST
\citep{WFIRST} are important for measuring cosmological parameters of
our Universe, for studying the evolution of galaxies, and for unveiling
the nature of dark matter and dark energy.  In order to capture as
much information as possible those survey observations have been increasing their
volume and redshift depth.  For example, the detection of the Baryonic Acoustic
Oscillations (BAO) in the distribution of Luminous Red Galaxies (LRG) in
the SDSS survey \citep{SDSSBAO} was based on 46,768 galaxies in a
 volume $0.72\,h^{-3}{\rm Gpc}^3$. The BOSS measurements of
cosmological parameters are based on 1.2 million LRGs in a
volume of $5.8\,h^{-3}{\rm Gpc}^3$  \citep{Alam2017}. The volume of the DESI/Euclid and LSST
surveys will be $\sim 50\,h^{-3}{\rm Gpc}^3$ and  $\sim 100\,h^{-3}{\rm Gpc}^3$ respectively.

Theory has responded to this enormous survey volumes by making
cosmological simulations of huge computational volumes with equally
increasing the number of dark matter particles and the supercomputing
resources. The Euclid Flagship Simulation, DarkSky and Outer Rim, with
more than one trillion particles in a volume of 5-8
$h^{-3}{\rm Gpc}^3$ on a side, are good examples of the state-of-the-art achievements made recently in this field
\citep{PKDGRAV3,DarkSky,Habib2016}.

\makeatletter{}\begin{table*}
\begin{center}
\caption{Numerical and cosmological parameters of different simulations.
  The columns give the simulation identifier, 
  the size of the simulated box in $h^{-1}\,{\rm Mpc}$,
  the number of particles, 
  the mass per simulation particle $m_p$ in units of $h^{-1}\,M_\odot$, the mesh size $\Ng^3$,
  the  gravitational softening length $\epsilon$ in units of $h^{-1}\,{\rm Mpc}$, the number of time-steps $N_s$, 
the amplitude of perturbations $\sigma_8$, and
  the number of realisations $N_r$. The last column gives references.}
\begin{tabular}{ l | r | c | l |   c | c |  c | r |r |r }
\hline  
Simulation & Box\phantom{1} & particles  & $m_p$\phantom{mmm}   & $\Ng^3$  & $\epsilon$ & $N_{\rm s}$ & $\sigma_8$  & $N_r$ & Refs.
\tabularnewline
  \hline 
A0.5         & 500$^3$    & 1200$^3$ & $6.16\times 10^9$   & 2400$^3$ & 0.208 & 181 & 0.822   & 680 & 1
\tabularnewline
A1          & 960$^3$   & 1200$^3$ & $4.46\times 10^{10}$ & 2400$^3$ & 0.400 & 136 & 0.822    & 2532 & 1
\tabularnewline
A1.5         & 1500$^3$   & 1200$^3$ & $1.66\times 10^{11}$ & 2400$^3$ & 0.625 & 136 & 0.822    & 4513 & 1
\tabularnewline
A2.5         & 2500$^3$   & 1000$^3$ & $1.33\times 10^{12}$ & 2000$^3$ & 1.250 & 136 & 0.822    & 1960 & 1
\tabularnewline
A2.5c         & 2500$^3$   & 1000$^3$ & $1.33\times 10^{12}$ & 2000$^3$ & 1.250 & 285 & 0.822    & 1600 & 1
\tabularnewline
C1.2       & 1200$^3$   & 1000$^3$ & $1.47\times 10^{11}$  & 3000$^3$ & 0.400 & 136 & 0.822  & 100 & 3
\tabularnewline
D0.25         & 250$^3$    & 1000$^3$ & $1.33\times 10^9$   & 2000$^3$ & 0.125 & 181 & 0.822   & 120 & 3
\tabularnewline
D2.75        & 2750$^3$   & 1100$^3$ & $1.33\times 10^{12}$ & 4400$^3$ & 0.6250 & 136 & 0.822    & 22 & 3
\tabularnewline
D4          & 4000$^3$   & 2000$^3$ & $6.82\times 10^{11}$  & 4000$^3$ & 1.000 & 136 & 0.822    & 100 & 3
\tabularnewline
MDPL          & 1000$^3$   & 3840$^3$ & $1.5 \times 10^{9}$  & --      & 0.010 & --  & 0.828     & 1 & 2
\tabularnewline
HMDPL          & 4000$^3$   & 3840$^3$ & $7.9 \times 10^{10}$  & --      & 0.025 & --  & 0.828     & 1 & 2
\tabularnewline
\hline
\multicolumn{10}{l}{\quad {\it References:} $^1$\citet{GLAM}, $^2$ \citet{Klypin2016} , $^3$ this paper }
\tabularnewline
\end{tabular}
\label{table:simtable}
\vspace{-5mm}
\end{center}
\end{table*}

 Effects of computational box size were the topic of extensive
  discussions for the last few decades with introduction of different ideas and
  presentation of numerical results \citep[e.g.,][]{Tormen1996,Cole1997,Klypin1996,
Jenkins1998,Tinker2008,Angulo2010,Klypin2016}.

In the modern field of large cosmological simulations it is taken for granted
that the ideal situation is when the volume of a single computational box covers
the whole effective volume of the observational survey
\citep[e.g.,][]{DarkSky,Comparat,PKDGRAV3,Habib2016}. But why is this true?  It
is clear why galaxy surveys must be large: we need to have as much
information as possible, and the only way to do it is to increase the
volume of the galaxy sample. However, what is the reason to have a 
single simulation box with a computational volume as large as possible? In the sense
of statistics of matter density fluctuations (and related abundance of halos, voids,
filaments and so on), one can produce as many realizations of the
``universe'' as needed in order to match the statistics seen in the observations. 
In terms of computational complexity (computational cost, access and
dissemination of the results) we are in a more comfortable situation with many smaller
simulation boxes. In any case we need to make many realizations to estimate
noises and covariances -- all needed for the data analysis of the surveys.

One can list a number of effects related with the finite volume of a
simulation box. Those include the impact of periodically replicated images when a small
computational box is replicated many times to mimic a large
observational survey, and the effect of missing long-waves on the halo mass
function, the clustering signal, and the covariance matrixes. Some of these
effects have been already discussed in the literature
\citep[e.g.,][]{Hu2003,Warren2006,DarkSky,SuperScale,GLAM}.  Here we
review the situation, and provide estimates and arguments, regarding the
effects of long-waves in cosmological large-scale structure simulations.

The starting issue here is what observable one wants to study. If
waves longer than $\sim 1$~Gpc are probed then there is no other
option but to mimic those waves in theoretical estimates by using
extreme computational volumes comparable to the size of the observable
universe. Examples of these type of observables are the measurements
of the power spectrum of fluctuations for wave-numbers
$k\lsim 0.001\kMpch$ or the two-point correlation function at $\sim 1\Gpch$ scale. In this case 
the computational volume must be extremely large.

However, in most of the cases the observables may not {\it explicitly}
involve extremely long-waves.  Consider as an example the abundance of
very massive ($\gsim 10^{15}M_\odot$) clusters of galaxies. Clusters
themselves have radii $\sim 2$\,Mpc and gather mass from
$\sim 10$\,Mpc regions around them. So, the clusters are relatively
small objects. However, their abundance depends {\it implicitly} on
longer waves because those waves non-linearly couple with $\sim 10$\,
Mpc waves, which are responsible for the formation of the clusters. Another
relevant example is the study of the Baryonic Acoustic Oscillations (BAO). The BAOs
manifest themselves as a peak in the correlation function at pair
separation of $\sim 100\Mpch$. Again, the signal of the peak is relatively small, but
may {\it implicitly} depend on very long-waves through non-linear
interactions.

The goal of this paper is to estimate the impact of missing
long-waves in finite volume simulations on some important statistics that 
depend implicitly on those long waves.

This paper is structured as follows. We give a short introduction in Section 1. In Section 2 we present the
suite of simulations used in this work. Methods and definitions are
discussed in Section 3, and the impact of box replication is described
in Section 4. The missing power estimates due to the lack of
long-waves in the computational simulation box are given in Section 5,
and the results of the impact on other statistics such as the
correlation function, PDF, halo abundances, power spectrum and
covariance matrix are presented on Sections 6, 7, 8 and 9. We study
Super Scale Covariances (SSC) in Section 10. Finally we conclude and
summarise our results in Section 11.

\section{Simulations}
\label{sec:sim}

Most of the results presented in this paper are based on cosmological
$N$-body simulations.  In Table~\ref{table:simtable} we present the
numerical parameters of our simulation suite: box-size, number of
particles, mass of a particle $m_p$, number of mesh points $N_g^3 $
(if relevant), cell-size of the density/force mesh $\epsilon$, the
number of time-steps $N_s$, cosmological parameters $\sigma_8$ and
$\Omega_m$, and number of realizations $N_r$.

Different codes were used to make those simulations. The MultiDark
Planck $1\,\Gpch$~ MDPL2 and $4\,\Gpch$ HMDPL simulations
\citep{Klypin2016} were done with the \textsc{gadget-2} code
\citep{Gadget2}. The other simulations were carried out with the
parallel Particle-Mesh code \GLAM\, \citep{GLAM}. Because the \GLAM\,
code is much faster than \textsc{gadget-2}, we have done many
realisations of the simulations with the same cosmological and
numerical parameters that only differ by initial random seed.  All the
\GLAM\ simulations were started at initial redshift $z_{\rm init}=100$
using the Zeldovich approximation.  These simulations span three
orders of magnitude in mass resolution, a factor of hundred in force
resolution, and differ by a factor of $10^5$ in effective volume. The
differences in box-size are large, which is important for analysis
done in this paper, i.e., from $L=250\Mpch$ to $L=4\Gpch$. We did not
study smaller boxes because simulations with $L\lsim 250\Mpch$ become
unpractical for large-scale structure studies even if finite box-size
effects were corrected. They would also require too much replication
to fill the observational volume . As we show below the box-size
effects become too severe in those small boxes for relevant statistics
such as the correlation function at the BAO peak and abundance of
clusters of galaxies.

All simulations and analytical results presented in this work use the same cosmological
parameters: a flat LCDM Planck cosmology with $\Omega_m=0.307$, $h=0.67$.

\section{Methods and definitions}
\label{sec:methods}

A finite box-size $L$ -- either in simulations or in analytical estimates --
yields an important parameter: the fundamental wavenumber, i.e.,

\begin{equation}
\kbox = \frac{2\pi}{L}.
\end{equation}
In order to estimate the matter power spectrum $P(k)$ from the \GLAM simulations we
generate the dark matter density field on a 3D-mesh of size $N_g^3$ (see Table~\ref{table:simtable}).  
The Cloud-In-Cell (CIC) density assignment is used to estimate the density
field. We then apply FFT to generate the amplitudes of $N_g^3$ Fourier
harmonics. The minimum spacing of the harmonics in phase-space is
$\Delta k = \kbox$. The power spectrum is obtain on a 1D-mesh with
constant binning equal to $\kbox$. Each harmonic contributes to two
mesh elements with the weights obtained using the CIC interpolation scheme in
the same fashion as that used for the density assignment \citep{GLAM}. This binning
procedure reduces the noise in the power spectrum by $\sim 30\%$.
The power spectrum is corrected for the aliasing due to the CIC density assignment.

The covariance matrix $C(k,k^\prime)$ of the power spectrum is defined
as a reduced cross product of the power spectra at different
wave-numbers $k$ and $k^\prime$ for the same realisation averaged over
different realisations:

\begin{equation}
C(k,k^\prime) = \langle P(k)P(k^\prime)\rangle - \langle P(k)\rangle \langle P(k^\prime)\rangle .
\label{eq:cov}
\end{equation}
The covariance matrix is typically normalized by the average amplitude of the diagonal
componentes and plotted as $[C(k,k^\prime)/P(k)P(k^\prime)]^{1/2}$.

When estimating the density distribution function (PDF) for a given simulation we use
a different 3D-mesh size $N$ not necessarily equal to the mesh size of the
simulation itself. The CIC density scheme is applied for every mesh
size used.  Once the overdensity field is created the values of
the overdensity $\rho = \rho_{\rm DM}/\langle \rho_{\rm DM} \rangle$
are binned using logarithmically spaced bins with width
$\Delta\log_{10}(\rho) =0.025-0.050$. The PDF is then defined as a
normalized number $\Delta N$ of cells with overdensity in the range
$[\rho,\rho+\Delta\rho]$, i. e.,
\begin{equation}
P(\rho) = \frac{\Delta N}{N^3\Delta\rho}.
\label{eq:PDF}
\end{equation}
By construction, the PDF is normalized to have the total volume and
the total mass density to unity.  The second moment of $P(\rho)$ is
the {\it rms} fluctuation of the overdensity field and is related to
the power spectrum of fluctuations in simulations by
\begin{equation}
\sigma^2 = \int^\infty_0(\rho-1)^2P(\rho)d\rho =
            \frac{1}{2\pi^2}\int_{\kbox}^{k_{\rm Ny}}P(k)W^2(k\Delta x)k^2dk,
\end{equation}
where $\Delta x = L/N$ is the cell-size of the density field and
$k{\rm Ny} = \pi/\Delta x$ is the Nyquist frequency of the mesh. Here
$W^2(k\Delta x)$ is the power spectrum of the CIC filter with cell
size $\Delta x$.

Depending on the cell size the PDF can have a very wide range of
values. For the relatively small cell-sizes $\Delta x = (1-5)\Mpch$ used
in this paper the leading term in the PDF is
$P(\rho)\propto \rho^{-2}$ \citep{Bouchet1991,PDF}. In order to reduce
the dynamical range of the PDF we typically plot $\rho^2P(\rho)$.

\section{Effects of box replications}
\label{sec:replicate}
\makeatletter{}\begin{figure}
\centering
\includegraphics[clip, trim= 1.5cm 5.cm 0.5cm 2.5cm, width=0.495\textwidth]{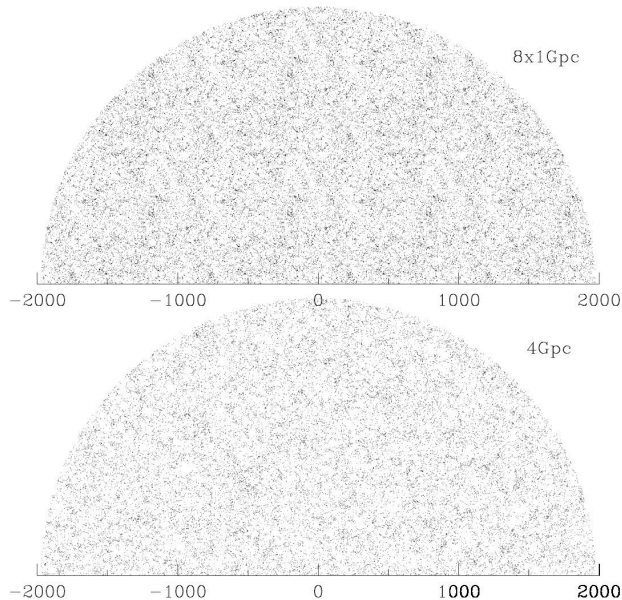}
\caption{Distribution of dark matter halos with mass
  $M>10^{14}\,h^{-1}M_\odot$ in a $400\Mpch$ slice with
  distances  $R<2000\Mpch$. The horizontal axis
  shows scales in $h^{-1}$~Mpc units. The bottom panel shows the sample distribution 
  for the HMDPL simulation with a computational box of $4\Gpch$ on a side. No
  replication was done in this case. The top panel is for the
  MDPL simulation with a computational box of $1\Gpch$ on a
  side. In this case the halo sample was periodically replicated many times to cover the
  same coordinate domain. This replication leads to repeating images
  of the same structures that can be seen more clearly in the central
  region of the plot.}
\label{fig:4gpc}
\end{figure}

If the computational box of the simulation is smaller than the volume
of a given galaxy survey, {\bf  the same} simulation box must be replicated
enough times to cover the entire observed region. Note that in order
to avoid defects at the boundaries of the box, the same realization
is replicated.  Box replications increase the apparent volume of the
sample as compared to the volume of a single simulation. However, they
do not add new information: it is still the same as in the original
simulation. For example, if long-waves were absent in the simulation
box, they will be absent in the replications. Nothing wrong with this:
it is understood that something will be missing if the replication is
applied to a finite volume simulation. The main question is: will the
replication procedure produce any defects?

One can imagine some possible issues. We start with the obvious one:
the same structure will be observed again and again due to the
periodical replication. Figure~\ref{fig:4gpc} illustrates the
situation. Here we use halos drawn from the MDPL ($1\Gpch$) and HMDPL
($4\Gpch$) simulations with virial masses larger than
$M>10^{14}h^{-1}M_\odot$.  We assume that in this case the
observational ``sample'' has a depth of $2000\Mpch$ and we also
show halos in a somewhat arbitrary chosen (but large) $400\Mpch$
slice. The bottom panel shows halos selected from the much larger
HMDPL simulation. No replication is needed in this case because the
HMDPL simulation covers the whole ``observed'' volume. The situation
is different in the case of the $1\Gpch$ MDPL simulation that requires
8 replications: 4 times along the x-axis and two along the
y-axis. Indeed one clearly sees the effects of the replications (see
top panel in Figure~\ref{fig:4gpc}) . This is obviously not a pleasant
feature: the real universe should not look like that. However, is it
really a problem? Once we agreed (or found) that waves longer than the
computational box are not important, then there is nothing wrong with
the top panel in Figure~\ref{fig:4gpc}. What we perceive as a defect
in the plot is just a way for our brain to tell us that there are no
waves longer than $1\Gpch$. Indeed, if we had analysed the new
(replicated) sample and ignored the effects of sample boundaries, we
would have found the same properties as in the original small volume
simulation -- the same halo abundances, peculiar velocities,
correlation function, and the same power spectrum truncated at the
fundamental mode of the simulation box.

\makeatletter{}\begin{figure}
\centering
\includegraphics[clip, trim= 1.5cm 5cm 0.5cm 2.5cm, width=0.49\textwidth]
{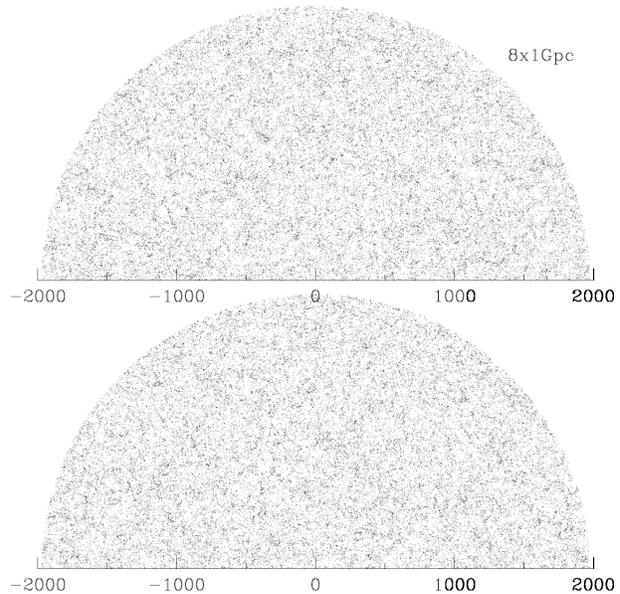}
\caption{The same as in Figure~\ref{fig:4gpc}, but with the
  computational volume of the $1h^{-1}$~Gpc MDPL simulation replicated
  and rotated along the x- and y- axes. The top and bottom panels are for
  two different rotation angles (see text). No periodical structures are seen in
  these images.}
\label{fig:4gpcrotate}
\end{figure}

There is a simple way to remedy the visual problems with the
replications.  One needs to rotate the stacked simulations before
making mock observational samples: the same realization is stacked and
the resulting distribution is rotated.  We illustrate this by rotating
twice the stacked distribution of halos in the $1\Gpch$ MDPL
simulation. We first rotate by some angle ($\sim 30^o-60^o$) the
distribution along the y-axis (the vertical axis in
Figure~\ref{fig:4gpc}), and then by another angle along the x-axis
(horizontal axis in the same plot). After the rotations are done, we
make the same slice as described before. Figure~\ref{fig:4gpcrotate}
shows two examples of mock samples produced in this way. The plots do
not show any visual defects of the replications. Just as in the case
of a simple replication, the rotated stacked distribution does not
bring new information. For example, if we estimate the power spectrum
of fluctuations of the rotated and stacked distribution, we will find
the same power spectrum as that found in the original $1\Gpch$ box
with shifted angles of the harmonics.

The other potential issue with the replication process is repeating structures
(halos, voids, filaments) along the line-of-sight. An example is the
study of the weak-lensing signal produced by clusters of galaxies or individual
galaxies. In order to mimic observations, the same simulation can be
repeated many times (stacked) along the line-of-sight with an
``observer'' placed on the line going through the centres of the
aligned boxes. If the box is small and the observer is at a large
distance from the lens, then every object will be found replicated
many times along the line-of-sight, which constitutes a serious defect
for the weak-lensing estimates.

\begin{figure} \centering
\includegraphics[clip, trim= 1.5cm 5cm 0.5cm 2.5cm, width=0.49\textwidth]
{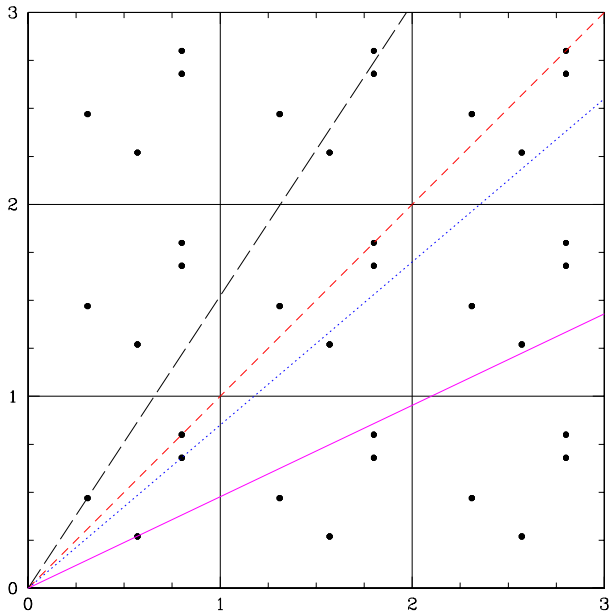}
\caption{Illustration of the effects of replication of a large
  simulation box. Four objects were placed in a square of size one and
  then replicated 3 times along the x-axis and three times along
  y-axis. The observer is placed in the left bottom corner of the
  resulting $3 \times 3$ square. The line-of-sight periodically
  replicated images that present a problem for the weak-lensing
  analysis are only for those objects along the diagonal of the square
  and its main axes, which are highly improbable configurations.}
\label{fig:replicate}
\end{figure}

The key issue here is the size of the simulation. If it is too small,
say 100--200\,Mpc, then indeed the replication is problematic. With
the typical distance to lenses of $\sim 1\Gpch$ a small
$\sim 100\Mpch$ box will result in almost plane-parallel projection on
the sky and, thus, with multiple halos almost exactly along the
line-of-sight. The situation is different for large simulations with
size $\sim 1\Gpch$.  In this case multiple replications are still
required, but they do not produce problems for weak-lensing estimates.
Figure~\ref{fig:replicate} schematically illustrates this situation
with the replications of a large computational box. To make the
problem more transparent we place four objects in a 2D-box of unit
size and replicate it 3 times in each direction. The ``observer'' is
placed in the corner of the box and the lines connecting the observer
going through each point are shown. Most of the lines do not have
periodical images. The only one that does is the object that is
exactly along the diagonal. We know which points will have periodical
images and which, thus, will have problems with lensing analysis. If
$(x, y)$ are the coordinates of the objects, then periodical images
will appear if the ratio of the coordinates is a rational number. In
other words, if $x/y = i/j$; where $i, j$ are integer numbers. A
periodical image appears after $i$ replications along the x-axis and
$j$ replications along the y-axis.  Because we replicate the
simulation box only few times (three for Figure~\ref{fig:replicate}),
we are potentially interested in the cases with small values of $i$
and $j$. In a mathematical sense the probability of an arbitrary $x$ and
$y$ to be a rational number is zero. In 3D the situation is even more
strict because two ratios $x/y$ and $y/z$ must be rational with small
integers. In practice the chance to have close images along the
line-of-sight of the same object are very small and can be found in
every case.

 \makeatletter{}

\section{Missing power}
\label{sec:missing}
\begin{figure}
\centering
\includegraphics[clip, trim= 0.5cm 5cm 0.5cm 2.5cm, width=0.49\textwidth]
{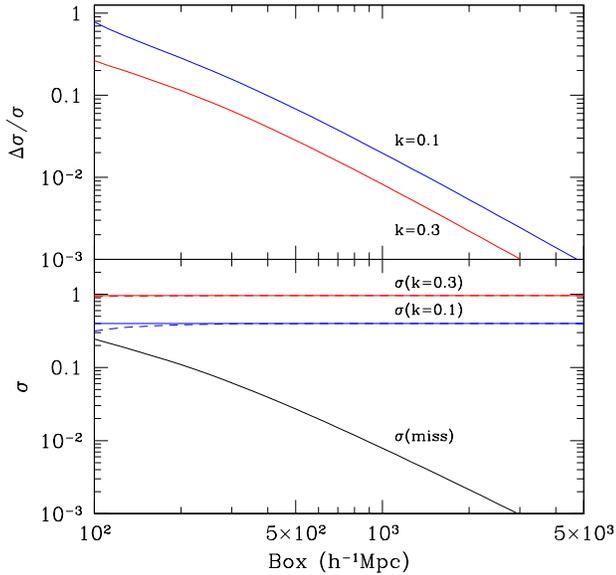}
\caption{Missing dark matter power of density fluctuations in
  simulations with different box sizes. {\it Bottom panel:} The lower
  full curve shows the $rms$ of density fluctuations in waves longer
  than the box-size $k<k_{\rm box}=2\pi/L$. The other curves show
  the $rms$ fluctuations up to $k=0.1h\Mpc^{-1}$ (lower curves) and
  $k=0.3h\Mpc^{-1}$ (top curves): the full curves show the total $rms$
  including waves up to the distance to the horizon while the dashed
  curves are for waves from the box size down to the specified
  wavenumber. {\it Top panel} shows the ratio of the missing power in
  waves with $k<k_{\rm box}=2\pi/L$ to the power inside the
  specified wavenumber indicated in the plot. The missing $rms$ power
  can be substantial for simulations with boxes smaller than
  $\sim 200\Mpch$. It becomes tiny for simulations with boxes
  larger than $\sim 1\Gpch$ where most of the missing power is in waves
  that are just a bit longer than the box-size. 
  }
\label{fig:sigma}
\end{figure}

The size of the computational volume defines another important
ingredient: the amplitude of the power missed in the simulation
box. The larger is the box the smaller is the missing power and, thus,
the simulation closely matches the density fluctuations in the
Universe.  We can estimate the missing power $\sigma_{\rm miss}$ by
integrating the linear power spectrum $P(k)$ from $k=0$ up to the
wavenumber given by the fundamental mode of the box $\kbox$, i.e.,
\begin{equation}
\sigma^2_{\rm miss}(L) =\frac{1}{2\pi^2}\int_0^{\kbox}P(k)k^2dk, \quad \kbox =2\pi/L.
\label{eq:miss}
\end{equation}
The missing power can be computed for any redshift, but here we will
do the estimates only for $z=0$.  The bottom panel in
Figure~\ref{fig:sigma} shows $\sigma_{\rm miss}(L)$ for different
box-sizes $L$.

The plot shows that the missing power declines dramatically with
increasing box-size. This is expected because at small $k$ the power
spectrum $P(k)$ is nearly primordial with slope $\sim 1$. Thus,
$\sigma^2 \propto k^4\propto L^{-4}$. While it is easy to estimate
$\sigma_{\rm miss}(L)$ numerically, it is convenient to have a simple
approximation for large simulation boxes and Planck cosmology:
\begin{equation}
  \sigma_{\rm miss}(L) \approx \frac{7.5\times 10^{-3}}{L^2_{\rm Gpc}},
 \quad L_{\rm Gpc}\equiv\frac{L}{1\Gpch}.
\label{eq:missapprox}
\end{equation}

The other side of this steep decline is that the missing power {\it
  increases} dramatically for small boxes. For example, for
$L=200\Mpch$ the missing power is $\sigma_{\rm miss}\approx 0.1$,
which is substantial considering that one expects that non-linear
effects (e.g., turn-around for halo formation) become important when
the overdensity becomes unity.  However, the missing power becomes
very small, and falls below $\sigma_{\rm miss}<10^{-2}$, for a
$L=1\Gpch$ simulation box.

There are different ways of assessing how large is the power missed in
a finite box size. The other lines in the bottom panel of
Figure~\ref{fig:sigma} correspond to the power in eq.(\ref{eq:miss})
integrated up to a giving wavenumber $k_{\rm cut}$ instead of
$\kbox$. We use two values of $k_{\rm cut}$: $0.1\kMpch$ and
$0.3\kMpch$ which are characteristic for the domain of the BAO
peaks. The full curves are for the total power (infinite box) and the
dashed curves are for the power inside the box (with the integrals
starting at $\kbox$). Clearly there is not much missing power except
for those boxes with $L\lsim 200\Mpch$. The top panel in
Figure~\ref{fig:sigma} shows the ratio of missing power in waves with
$k<k_{\rm box}$ to the power inside the specified wavenumber indicated
in the plot. The missing $rms$ power can be substantial for
simulations with boxes smaller than $\sim 200\Mpch$, but it becomes
tiny for simulations with boxes larger than $\sim 1\Gpch$.

\makeatletter{}

It is also interesting to note that most of the missing power
$\sigma_{\rm miss}(L)$ is found in waves that are just a bit longer
than the computational box. For example, for a $1\Gpch$ box 95\% of
the missing power is in waves with wavelengths between $(1-2)\Gpch$
and 88\% is in $(1-1.5)\Gpch$ waves. These waves cannot be considered
constant inside the computational box: a striking contrast with the
main presumption of the separate universe simulations
\citep[e.g.,][]{SuperScale,Wagner2015} which assumes that the only
long-waves that matter are those that are much longer than the length
of the computational box, and, thus, can be treated as a constant
background. The $rms$ density fluctuation $\sigma_L$ of the average
density inside a box $L$ embedded in an infinite density field is
about 5 times smaller than $\sigma_{\rm miss}(L)$. See
Section~\ref{sec:SSC} for more details.

\makeatletter{}\begin{figure}
\centering
\includegraphics[clip, trim= 1.5cm 5cm 0.5cm 2.5cm, width=0.49\textwidth]
{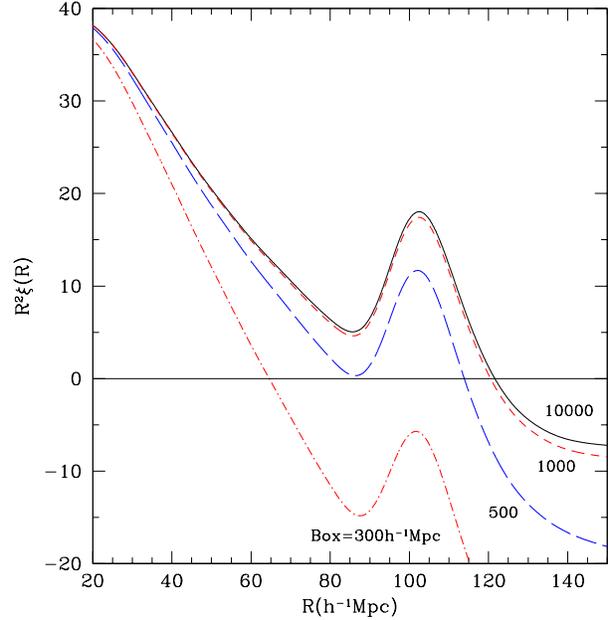}
\caption{Effects of the box-size on the correlation function $\xi(R)$
  in the linear regime. We estimate the correlation function using the
  linear power spectrum truncated on wavenumbers smaller than the
  fundamental mode, $k< k_{\rm box}$, with the box-size indicated in
  the plot. The peak of $\xi$ at $R_{\rm BAO}\approx 100\Mpch$ is due
  to the BAO. For larger box-sizes the correlation function crosses
  zero at $R_0\approx 122\Mpch$. When the box-size becomes small the
  amplitude of the correlation function decreases at large radii:
  the zero-crossing shifts to much smaller distances and the BAO
  amplitude is severely affected.}
\label{fig:correlation}
\end{figure}

\section{Impact on the correlation function}
\label{sec:corr}
\makeatletter{}\begin{figure}
\centering
\includegraphics[clip, trim= 0.5cm 5cm 0.5cm 2.5cm, width=0.49\textwidth]
{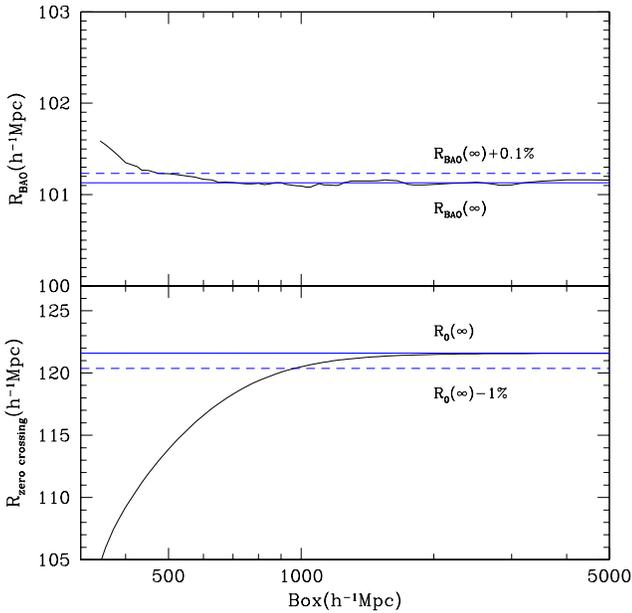}
\caption{Dependance of the BAO peak $R_{\rm BAO}$ (top panel) and the
  zero-crossing $R_0$ (bottom panel) on the box-size in the linear
  correlation function. The position of the BAO peak is very
  insensitive to the box-size with 0.1\% change for a box with
  $L \approx 500\Mpch$. The zero-crossing $R_0$ is much more
  affected: for better than 1\% error the simulation box must be
  larger than $1\Gpch$.}
\label{fig:corrMax}
\end{figure}

Because of the truncation of the power spectrum at the fundamental
mode, the finite-size box correlation function of the dark matter is
different at large scales from that expected when one assumes an
infinite volume \citep{Sirko2005,Klypin2013}. The correlation
function of the dark matter or that of halos are affected by
non-linear processes. Still, their main features (e. g., position of
the BAO peak and zero-crossing; see Figure 5 in \citet{Klypin2013})
are reproduced by the linear theory with some modifications though. In
any case, it is important to estimate how accurately we can even
reproduce the linear correlation function.

  The finite box-size correlation function $\xi(R)$ can be estimated
  using the power spectrum $P(k)$, i. e.,
\begin{equation}
\xi(R) = \frac{1}{2\pi^2}\int_{\kbox}^\infty dk k^2 P(k)\frac{\sin(kR)}{kR}.
\end{equation}
Figure~\ref{fig:correlation} presents the estimates of the correlation
function of the linear dark matter power spectrum for different box-sizes.  
Just as expected, the differences become small at smaller
scales. Indeed, the comparison of the correlation functions of halos
in the Bolshoi ($L=250\Mpch$) and MultiDark ($L=1\Gpch$) simulations
are also within few percent for $R<10\Mpch$ \citep{Klypin2013}.

At larger scales the box-size effects become more apparent. For
example, for a $L=300\Mpch$ box the correlation function is
qualitatively incorrect: the whole BAO domain is negative and the
zero-crossing scale is twice smaller than it should be (see
Figure~\ref{fig:correlation}).  The situation
improves when the box-size increases. However, the box-size should be
substantially larger than $500\Mpch$ in order to closely match the
correlation function of the infinite box.

Just as with the estimates of the missing power, the effects due to
the missing long-waves dramatically decline with increasing of the box
size. Indeed, we can hardly see any impact for $L=1\Gpch$. We can
quantify the effect using two statistics: the position of the BAO peak
$R_{\rm BAO}$ and the scale of zero-crossing $R_0$.  These two
parameters are plotted in Figure~\ref{fig:corrMax}. As we can see, the
position of the BAO is remarkably stable. For the $L=500\Mpch$ box the
BAO peak is within 0.1\% from its pristine location, and the
deviations become unmeasurable for larger boxes. This is good news
because the BAO position is an important parameter for estimates of
the cosmological parameters. It will be modified by non-linear
effects, but at least we start with an accurate linear theory
position.

The zero-crossing is much more sensitive to the box-size with large
uncertainties for boxes with $L<500\Mpch$. Still, the error decreases quickly with
increasing the box-size, and becomes less than 1\% for $L>1\Gpch$.

\makeatletter{}\begin{figure}
\centering
\includegraphics[clip, trim= 1.1cm 5cm 0.5cm 2.5cm, width=0.49\textwidth]
{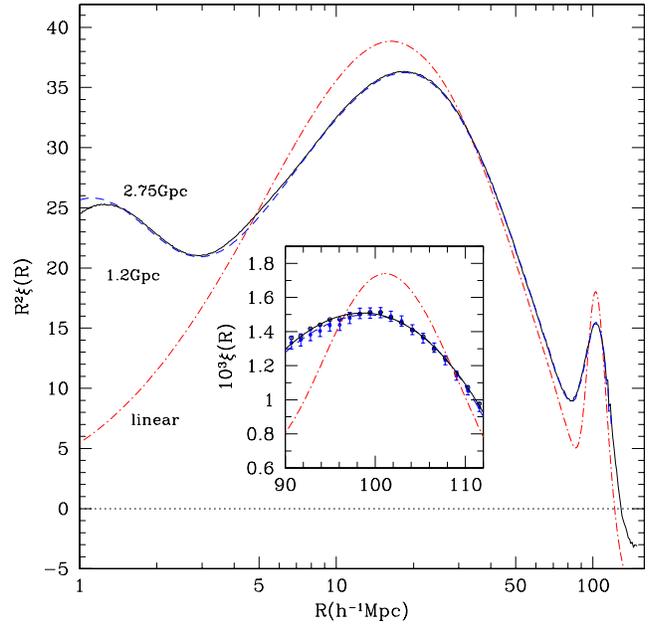}
\caption{Comparison of nonlinear dark matter correlation
  functions. Full and dashed curves show the results from the D2.75
  and C1.2 GLAM simulations correspondingly.  Differences at
  $R<5\Mpch$ radii are explained by the higher resolution of the C1.2
  simulations. At larger scales there are no measurable differences
  between the $L=1.2\Gpch$ and the much larger $L=2.75\Gpch$
  simulation boxes.  The insert in the figure shows in more detail the
  region around the BAO peak.  The error bars in the plot correspond to
  $1\sigma$-errors of the mean as evaluated using the 100 realisations
  of the C1.2 simulation. Full and dashed curves present analytical
  fits eq.(\ref{eq:xifit}) with differences in the position of the BAO peak less than
  0.1\%.}  

\label{fig:nonlincor}
\end{figure}

We study also the effects of nonlinear evolution using the C1.2 and
D2.75 GLAM simulations at $z=0$. Figure~\ref{fig:nonlincor} presents
the average correlation function of dark matter in these simulations
for a wide range of radii $R=(1-150)\Mpch$. If the long-waves missed
in the C1.2 simulation boxes, as compared with the much larger boxes
of the D2.75 simulations were important, we would have seen a stronger
clustering in 2.75\Gpch box simulations at all scales. However, this
does not happen: there are no measurable differences between the
1.2\Gpch\, and the much larger 2.75\Gpch\, simulations for scales
$R>5\Mpch$.

In order to quantify the differences in the BAO domain, we fit the
average correlation functions of each set of simulations with an
analytical function -- a third order polynomial in the form:

\begin{equation}
\xi_{\rm fit}(R) = \xi_0+a_1x+a_2x^2+a_3x^3, \quad x\equiv R-R_0.
\label{eq:xifit}
\end{equation}
 The function has 5 free parameters with $R_0$ and $\xi_0$ defining the
position and amplitude of the peak of the correlation function. After
fitting the data in the range of radii $R=(91-113)\Mpch$ we find for
$L=2.75\Gpch$ simulations $\xi_0=1.500$, $R_0=100.30\Mpch$, which is
nearly identical (within 0.06\% for $R_0$) with those for the C1.2
simulations: $\xi_0=1.494\pm 0.005$, $R_0=(100.24\pm 0.1)\Mpch$.

The only statistically significant differences between D2.75 and C1.2
correlation functions are observed at small radii $R<5\Mpch$, which
are due to the differences in the force resolution.  This indicates
that the $1.2\Gpch$ box of the C1.2 simulations is large enough to
produce accurate results for the scales presented in
Figure~\ref{fig:nonlincor}.

\section{Density distribution function}
\label{sec:PDF}

The density distribution function of the dark matter $P(\rho)$
provides an additional test for the effects of the finite-box size $L$. One
may expect some impact due to the missing waves. Indeed, a very long wave with
a wavelength longer than $L$ increases the $rms$ fluctuations inside
the computational box. As the result, some fluctuations collapse
earlier when the density of the universe is larger. Thus, the
collapsed density will be somewhat larger as compared with the
situation when the long-wave is missed in simulations with box-size
$L$. Using the same argument, one expects that some regions will have
lower density, if the long-wave is present. In other words, the
density distribution function should be wider in simulations with larger boxes.
This is the same type of arguments that were mentioned in the
estimates of the halo abundances: the effect must be present, but
how large is it?

Here we will be interested in the high-density tail of $P(\rho)$
because of two reasons: (1) the power spectra and correlation
functions -- being averages over the whole computational volume --
have already gave us results on the properties of the density
field. However, they may not be very sensitive to a small fraction of the
volume with the largest density; (2) Because of the particle noise in regions
with low-density, it is more difficult to reliably estimate the PDF at low $\rho$.

We select three GLAM simulation sets to study the PDF. The main
comparison is between D2.75 and A1.5 with $L=2.75\Gpch$ and
$L=1.5\Gpch$), which have almost two times different box-sizes and
the same force resolution. So, the difference between those
simulations at large densities should be only due to the
box-sizes. However, these simulations have almost ten times different
number densities of particles that affects the low-density part of
$P(\rho)$. In addition, we also consider the A2.5 simulations that
have the same number-density of particles as D2.75, nearly the same
volume, but twice lower resolution. The density distribution function
$P(\rho)$ is estimated for three filtering scales -- sizes of cubic
cells: $\Delta x = 1.25, 2.5, 5.0 \Mpch$.

  \makeatletter{}\begin{figure} \centering
\includegraphics[clip, trim= 1.2cm 5cm 0.5cm 2.5cm, width=0.49\textwidth]
{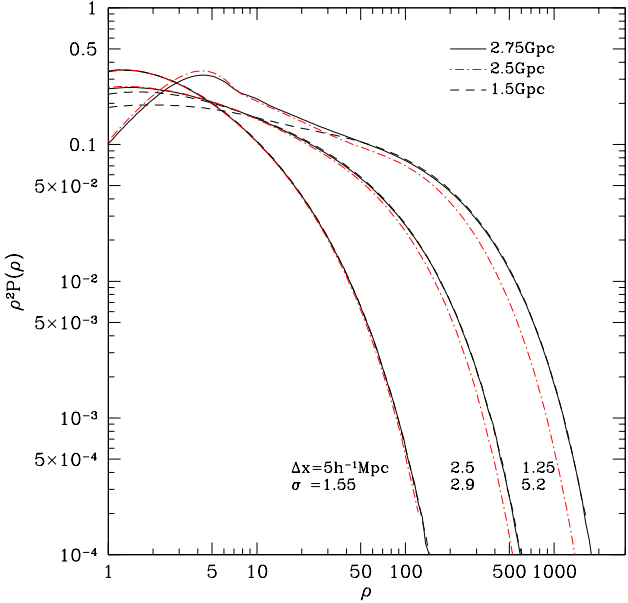}
\caption{The dark matter density distribution function $P(\rho)$ at
  $z=0$ for simulations with different box and
  cell-sizes. The density is given in the units
  of the average density of the Universe. The PDF is scaled with the
  square of density to reduce the dynamical range. The $rms$ density
  fluctuation $\sigma$ measured for different cell-sizes is indicated
  in the plot. 
  The lack of force resolution in A2.5
  results in the decline of PDF at large densities $\rho> 100$, while
  the particle noise becomes important for low densities $\rho <10$.
  In the regime where both the force and mass resolutions are small
  the PDF does not show any signs of dependance on the size of
  simulation box.
  }
\label{fig:pdf}
\end{figure}

The dark matter density distribution functions $P(\rho)$ are shown in
Figure~\ref{fig:pdf}. The density is given in the units of the average
density of the Universe. The PDF is scaled with the square of density
to reduce the dynamical range. The $rms$ density fluctuation $\sigma$
measured for the different cell-sizes is indicated in the plot. For the
large cell-size $\Delta x=5h^{-1}$Mpc the PDFs of the different box sizes
are practically indistinguishable. As the cell-size decreases, the
lack of the force resolution in the A2.5 simulations results in the decline of the
PDF at large densities $\rho> 100$, while the particle noise becomes
important for low densities $\rho < 10$.  In the regime where both the
force and mass resolutions are small the PDF does not show any signs of
dependance on the size of the simulation box.

\makeatletter{}\begin{figure} \centering
\includegraphics[clip, trim= 0.5cm 5cm 0.5cm 2.5cm, width=0.49\textwidth]
{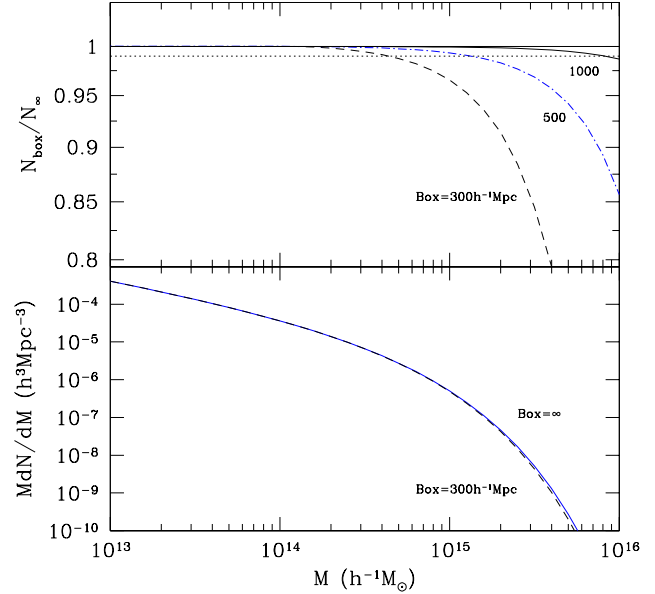}
\caption{Analytical estimates of the abundance of dark matter halos
  assuming a truncated linear power spectrum at long waves. The bottom panel shows
  the halo abundance for a spectrum truncated at the fundamental mode
  of a $300\Mpch$ simulation box (dashed curve) as compared to that obtain with an
  untruncated spectrum (full curve). The top panel shows the ratio of the
  predicted halo abundances in simulations with different box-sizes to
  that assuming an untruncated spectrum. The dashed line shows 1\%
  decrease in the halo abundance. There is almost no effect for halos
  with mass less than $10^{14}\Msunh$. The abundance of most massive
  cluster-size halos with $M\approx 10^{15}\Msunh$ can be underestimated by
  $\sim 5\%$ in simulations with a box-size of $300\Mpch$, but the effect
  dramatically decreases with increasing the simulation size and
  no measurable effect is observed for a $\sim 1\Gpch$ box.}
\label{fig:abundance}
\end{figure}
\section{Halo abundances}
\label{sec:abundance}
Missing large-scale power in finite box simulations must affect the
estimates of the abundance of halos with different masses. All
current analytical models -- build and tested using $N$-body results --
tell us that halo abundance is a function of the $rms$ density fluctuations
$\sigma(M,z)$ as  estimated from the linear power spectrum smoothed
with filter of effective mass $M$ at redshift $z$. Because the finite
box simulations miss some fraction of $\sigma$ for a given mass $M$,
these simulations must predict fewer halos.

However, so far the results provided by $N$-body cosmological
simulations have failed to show that this is the case
\citep{Warren2006,Tinker2008,DarkSky,Ishiyama2015,DeRose}. The halo mass
functions estimated using simulations of different box sizes have been
extensively studied in the field. For example, \citet{Tinker2008} used
simulations with sizes $L=80\Mpch$ up to $L=1.3\Gpch$.
\citet{DarkSky} analysed simulations with different box sizes between
$100\Mpch$ and $8\Gpch$. None of those works indicated any dependance of
the halo mass function on the simulation box-size.

In the overlapping halo mass interval
$M=10^{13}-2\times 10^{14}\Msunh$ the DarkSky simulations with boxes
$L=0.8, 1.6, 8\Gpch$ have mass functions that deviate by less than 1\%.
\citet{DeRose} do not find any differences
in the halo mass function of halos more massive than $\sim 10^{13}\Msunh$
when comparing simulations with $1\Gpch$ and $5\Gpch$ boxes. 
\label{sec:power}
 \makeatletter{}\begin{figure}
  \centering
\includegraphics[clip, trim= 1.5cm 5cm 0.5cm 2.5cm,width=0.49\textwidth]
{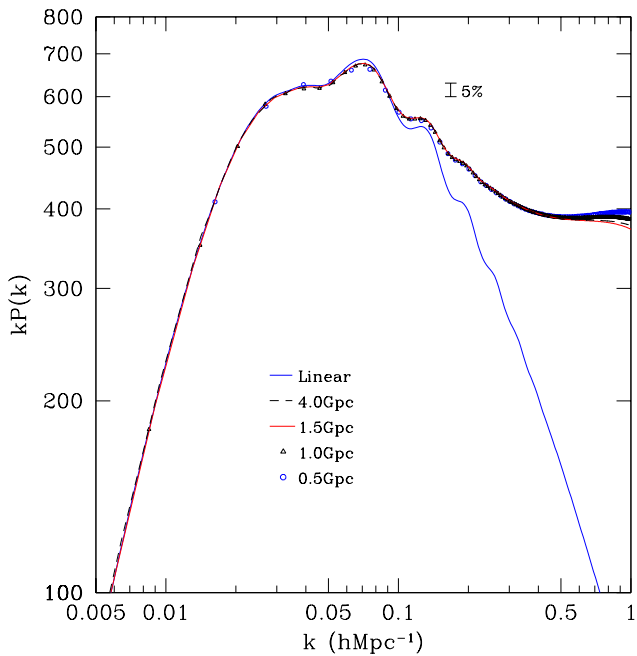}
\caption{Power spectrum of dark matter fluctuations at $z=0$
  scaled with wavenumber $k$. Simulations with different computational
  box-sizes are labeled with different symbols as indicated in the
  plot. Differences between simulations  at large $k\sim 1\kMpch$ are related with
  the force resolution. On larger scales (small
  $k<0.1h\Mpc^{-1}$), where the force resolution is not important,
  there are no visible signs  that the box-size affects  the power spectrum.}
\label{fig:power}
\end{figure}

In order to interpret and understand this result we use the analytical
approximation of the halo mass function $n(M) = f(\sigma(M,z))$, where
$\sigma(M,z)$ is the $rms$ of density fluctuations as presented in
\citet{Comparat}. By
itself this approximation is based on the MultiDark \citep{Klypin2016}
suite of simulations with box sizes $L=0.4-4\Gpch$.
We use this approximation to find the halo mass function in two
ways. First,  we estimate the $rms$ of fluctuations $\sigma(M,z)$
using the full (untruncated) linear power spectrum of
fluctuations. Second, we mimic the finite box-size effects by
truncating the power spectrum at the fundamental mode $\kbox$.

Figure~\ref{fig:abundance} presents our estimates of the halo mass
function for three hypothetical simulations with box sizes $L=300, 500\Mpch$ and
$1\Gpch$. Clearly one should expect some deficit of halos in the
simulation with $L=300\Mpch$. For example, for mass
$M= 2\times 10^{15}\Msunh$ the model predicts that about 10\% of the
halos will be missed. It is also clear why this effect has not been
measured in the $N-$body simulations, and why it could be ignored: the
model predicts that one should find only about one cluster for this
halo mass in such small box. Note that when analyzing the
simulations, one routinely ignores the first $\sim 100$ most massive halos
because these clusters are too sensitive to cosmic variance and also
because of the large statistical errors. If we limit ourselves to a
mass scale with more than 100 halos, then the $L=300\Mpch$ box
yields less than 1\% uncertainty in the halo mass function at the most
massive tail.

Figure~\ref{fig:abundance} also shows that the finite box-size
uncertainties decline dramatically with $L$. A simulation with a
box-size of $1\Gpch$ will end up with no missing clusters: 1\% error
is reached for a halo mass of $\sim 8\times 10^{15}\Msunh$. The
predicted number of clusters with this mass is so low that
no single cluster of this mass is expected in the Universe.

So, when it comes to making a choice for the box-size, our selection
depends on the observational sample, i. e, how massive are
the clusters in the sample that will be analyzed.  For example, if the
observed volume is relatively small and we are focusing only on
clusters with mass less than $10^{14}\Msunh$ then even a $300\Mpch$
box-size would be sufficient: on average it will produce the correct
amount of clusters under consideration. If instead we deal with a very
large survey and study all possible clusters, then the box-size must
be not less than $1\Gpch$.

The errorbars in the observed number of objects, which we estimate
using simulations, are the sum of two factors: (1) the statistical fluctuations
due to the cosmic variance (random noise due to all harmonics with
wavelength less than $L$) and (2) the effects of waves longer than the
computational box. The first term will be found by measuring the
statistics of objects in many realizations of simulated boxes.

The second term can be estimated by assuming that the number of
objects $n(M)$ depends on the $rms$ of density fluctuations at a given
scale $\sigma(M,z)$.  We use the halo mass function as an example. The
number of missed halos $\Delta n(M)$ due to change in $\sigma$ can be
written as
\begin{equation} 
\frac{\Delta n}{n} = \frac{\partial \ln n(\sigma)}{\partial \sigma} \Delta\sigma,
\end{equation}
where $\Delta\sigma= \sigma_{\rm miss}(L)$ is the $rms$ fluctuations
due to waves longer than $L$. Note that this is exactly the quantity
that is plotted in the top panel of Figure~\ref{fig:abundance}.  The
errors can be substantial for $\sim 10^{15}\Msunh$ clusters in
$300\Mpch$ simulations, but they are negligible for any clusters in
$1\Gpch$ runs.

\section{Finite-box effects on the Power spectrum}

\makeatletter{}\begin{figure}
\centering
\includegraphics[clip, trim= 1.5cm 5cm 0.5cm 2.5cm, width=0.49\textwidth]
{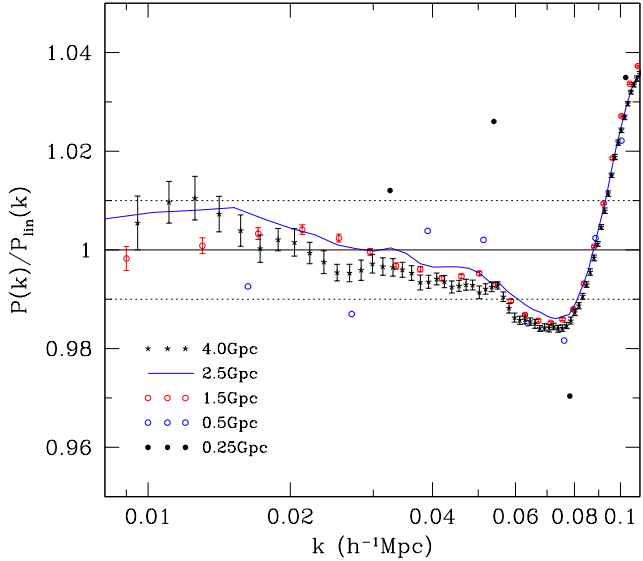}
\caption{The ratio of the non-linear to linear power spectrum --
  square of the bias parameter -- for simulations with different
  box-sizes. The bias parameter is scale dependent on
  $k\gsim 0.05h\Mpc^{-1}$. To avoid clutter we show statistical errors
  only for D4.0 and C1.5 simulations.  Because of the small number of
  realizations the $4\Gpch$ simulations show relatively large
  noise. Small-box simulations (D0.25 and D0.5) have $\sim 2$\%
  deviations from the average trend due to noise. Simulations with box
  sizes $>1\Gpch$ do not indicate a trend with the box-size.}
\label{fig:bias}
\end{figure}
The accuracy of the non-linear dark matter power spectrum from
$N$-body simulations has been addressed extensively in many
publications
\citep[e.g.,][]{Heitmann2008,Heitmann2010,Schneider2016,Lawrence2017,Smith2018}.
However, typically the main focus of these works is devoted to the
convergence of the results on the short-scales (see also
\citet{Smith2018}). Unlike the short scales, where the comparison of
just one or few realisations with different box-sizes is sufficient,
the analysis of the power spectrum for long-waves ($k\lsim 0.3\kMpch$)
is complicated due to the large cosmic variance, which would require
many realisations in order to complete a detailed study.
\citet{Heitmann2010} compared the power spectrum results obtained with
$234\Mpch,\, 960\Mpch$ and $2\Gpch$ simulation boxes. They find that
the power spectrum of 137 realisations of the $234\Mpch$ boxes is
below the larger box simulations by about $\sim 1\%$ for wavenumbers
$k=0.03-0.15\kMpch$. There were no detectable differences (less than
$\sim 1\%$ between their $960\Mpch$ and $2\Gpch$ boxes). \citet{GLAM}
used thousands of realisations to study the effects of the simulation
box-size on the average of the power spectrum. Here, we extend that
analysis to study in detail the effects of longer waves with
additional simulations.

Similar to the situation with the halo abundance, we know how
  qualitatively the missed long waves affect the power spectrum: power
  spectrum must increase with increasing of the box size. The
  magnitude of the effect is difficult to estimate. However, we know
  that the missing power is small for any realistic box-size (see
  Figure~\ref{fig:sigma}). Thus, most of the effect is expected to be
  found on long-waves in a given computational box. However, these
waves are still in nearly linear regime and their non-linear coupling
with the small amplitude waves outside the box can be expected to be
small.

\makeatletter{}\begin{figure}
\centering
\includegraphics[clip, trim= 1.5cm 5cm 0.5cm 2.5cm, width=0.49\textwidth]
{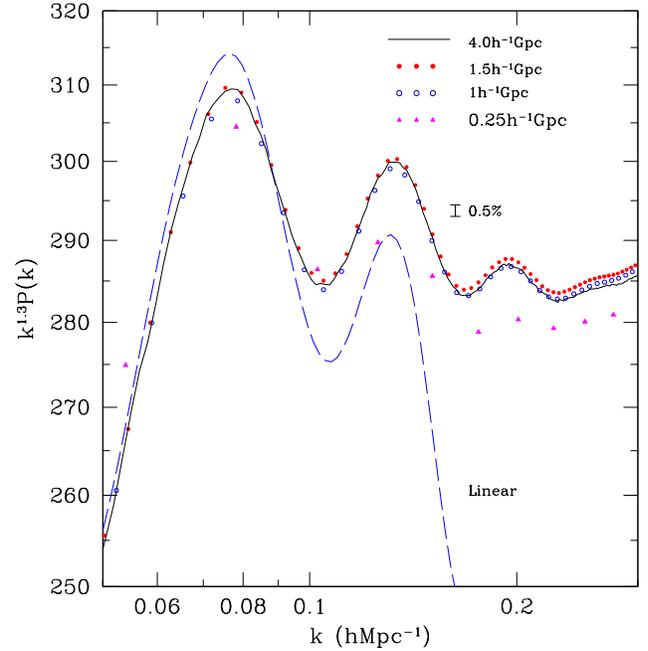}
\caption{Power spectra of dark matter in the domain of BAO peaks. The
  spectra are scaled with a $k^{1.3}$ factor to reduce the dynamical
  range. The vertical bar in the plot corresponds to 1/2 of percent
  deviations. The large-box simulations with $L\gsim 1\Gpch$ show
  remarkable degree of convergence with the differences less than
  $\sim 0.1\%$. The small-box simulations D0.25 systematically fall
  below the rest by $\sim 1-1.5$\%.  }
\label{fig:powerBAO}
\end{figure}

The average power spectra obtained from the different sets of
simulations are shown in Figure~\ref{fig:power}. The only differences
one can see in this plot are those due to the force resolution:
increasing resolution in small-box simulations results in the
  increase of the amplitude of the power spectrum. This happens at
large wavenumbers $k\gsim 0.5\kMpch$, which is a clear signature of
the resolution effects. The finite-box effects should act in the
opposite direction by decreasing the power in small-boxes relative to
the bigger ones. There are no obvious signs of the box-size effects on
long-waves where one expects them to be present.

In order to see the effects on small wavenumbers more clearly, we plot
the ratio of the nonlinear power spectrum to the linear spectrum
$P(k)/P_{\rm linear}(k)$ -- the square of the bias parameter. The
results presented in Figure~\ref{fig:bias} do not show any signatures
of depression in the power spectrum due to the missing long-waves. The
outliers in this panel are coming from the small $L=250\Mpch$ and
$L=500\Mpch$ simulations. There may exist a small effect of the
box-size at $k=(0.01-0.02)\kMpch$ where the bias systematically
increases by $\sim 0.5\%$ with increasing box-size, but the deviations
are within the statistical uncertainties due to the small number of
realisations of the $4.0\Gpch$ box.

One effect is nevertheless noticeable: the large spacing between
  points for small-box simulations. This is related with the
  fundamental harmonic $\kbox$ that defines discreteness effects in
  the Fourier space (minimum separation of harmonics): the larger is
  the box, the smaller is the binning. This can be a serious problem
  for small boxes. For example, for $L=250\Mpch$ the minimum  width
  of a bin $\Delta k$ is $\Delta k=0.025\kMpch$, which should be
compared with the wavenumber of the first BAO peak $\sim 0.07\kMpch$.
So, the binning is smaller than the BAO wavenumber, but only $\sim 3$
times. Indeed, the points in Figure~\ref{fig:bias} that deviate from
the other estimates of the $P(k)/P_{\rm linear}(k)$ ratio are those
that correspond to the small $L=250\Mpch$ simulations.

Figure~\ref{fig:powerBAO} presents a zoom-in view on the BAO domain of
the power spectrum. We multiply the power spectrum $P(k)$ by factor
$k^{1.3}$ with the goal to flatten the curves in the range
$k=(0.1-0.3)\kMpch$.  Simulations D0.25 with small box sizes clearly
suffer from the lack of long-waves: their power spectrum is
systematically fall below the rest by $\sim 1-1.5$\%. This is
consistent with estimates of \citet{Heitmann2010}.  There are no
measurable deviations between $1\Gpch$ and $4\Gpch$ simulation boxes
with differences less than $\sim 0.1\%$.

\makeatletter{}\begin{figure}
\centering
\includegraphics[clip, trim= 1.5cm 5cm 0.5cm 2.5cm, width=0.49\textwidth]
{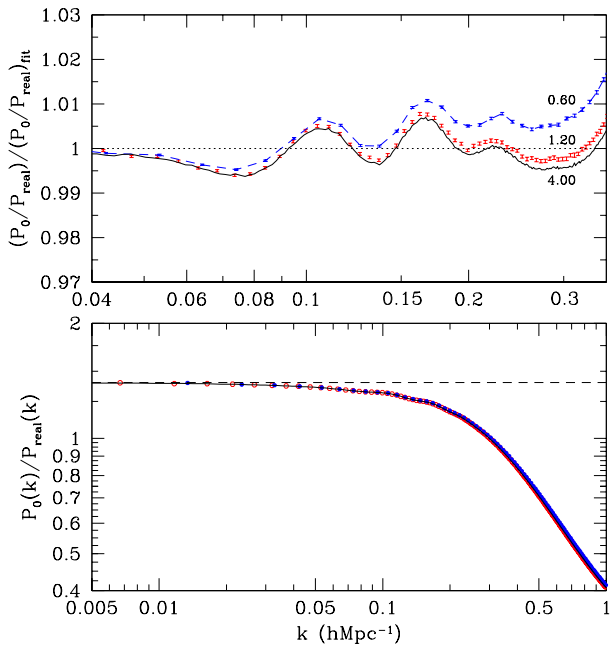}
\caption{ Redshift distortions in simulations with different box
  sizes. Full curves show results for D4 simulations with 4~Gpc
  box. These are compared with results for much smaller simulations
  with 1.2~\Gpch~ and 600\Mpch~ box sizes. {\it The bottom panel:}
  ratio of the redshift-space dipole power spectrum $P_0$ to the
  real-space $P_{\rm real}$. The horizontal dashed line indicates
  theoretical prediction for very long waves \citep{Kaiser1987}. Open
  (red) and filled circles (blue) are for simulations with the same
  mass and force resolution as D4 but in 1.2~\Gpch~ and 600\Mpch~
  boxes correspondingly. {\it The top panel:} Deviations
  $P_0/P_{\rm real}$ ratios from a smooth analytical function
  eq.(\ref{eq:fit}).  Differences between 4~Gpc and 1.2~Gpc are very
  small $\lsim 0.1\%$ at all scales.  Decreasing the box size to
  $600\Mpch$ results in $\sim 1\%$ errors at $k\gsim 0.2\kMpch$ and no
  measurable errors at very long waves $k\lsim 0.05\kMpch$.}
\label{fig:powerRSD}
\end{figure}
Presented results so far were done for quantities defined in real
space and did not include peculiar velocities.  The latter produce
distortions in the redshifts space that are important component of the
interpreting and understanding of observed clustering of objects
\citep[e.g.,][]{Kaiser1987,Hamilton1998,Reid2011,Sanchez2017}. Because
our results are based on $N$-body simulations where density and
velocity perturbations play equally important roles, accurate
estimates of the growth and evolution of density fluctuations imply
accurate estimates of peculiar velocities. In other words, convergence
of different statistics of density distribution guarantees convergence
of quantities in redshift space. To make this argument more clear, we
compare redshift space power spectra with those done in the real
space.

  When estimating the redshift-space power spectra, we perturb
  positions of particles along one of coordinate axes according to their
  peculiar velocities and periodically wrap them around, if necessary.
  Once the density in the redshift space is constructed, we find the
  spectrum and estimate either the monopole or quadrupole power
  spectrum. Results are averaged over three directions of velocity
  distortions.

  We use 100 realizations of the D4 simulations to make estimates of
  power spectra for large-box simulations. To find effects of the box
  size we additionally made 100 realizations with a $1.2\Gpch$
  box-size and 800 realizations with twice smaller $600\Mpch$ boxes. For these
  simulations we use exactly the same mass and force resolution as for
  the D4 simulations: $600^3$ particles moving in a $1200^3$ mesh for
  $1.2\Gpch$ simulations and $300^3$ particles moving in a $600^3$ mesh for
  $600\Gpch$ simulations. Because we are interested only in the
  effects of peculiar velocities, we analyze the ratios of the
  redshift to real-space power spectra. This greatly reduces the
  cosmic variance and allows us to dramatically reduce the statistical
  noise.

  Figure~\ref{fig:powerRSD} presents results for the monopole
  component (quadrupole component shows similar results). Full curves
  show results for the $4\Gpch$ box. Different symbols are for the
  $600\Mpch$ and $1.2\Gpch$ boxes. The bottom panel shows the ratio of
  the redshift-space dipole power spectrum $P_0$ to the real-space
  $P_{\rm real}$. The horizontal dashed line indicates theoretical
  prediction for very long waves \citep{Kaiser1987}:
  $P_0=(1+2f/3+f^2/5)P_{\rm real}$, where $f=d\ln\Delta/d\ln a$ is the
  grows rate of linear waves \citep[e.g.,][]{Reid2011}. Differences between
  simulations with different box sizes are so small that it is
  difficult to see them.  To find the differences, we fit the ratio
  $P_0/P_{\rm real}$ with an analytical smooth function and display
  the deviations from the same fit in the top panel.  The function
  itself is motivated by approximations used in the
  field. Specifically we use analytical approximation:
\begin{equation}
  \left(\frac{P_0}{P_{\rm real}}\right)_{\rm fit}=A\exp\left[-\left(\frac{k}{k_0}\right)-\left(\frac{k}{k_1}\right)^2\right],
\label{eq:fit}
\end{equation}
where $A=1.40$, $k_0=2.4h{\rm Mpc}^{-1}$, and
$k_1=0.66h{\rm Mpc}^{-1}$. The differences between $4\Gpch$ and
$1.2\Gpch$ simulations are small: less than $\sim 0.1\%$ on all
scales. At even smaller $L=600\Mpch$ simulations show some differences
at $k\gsim 0.1\kMpch$. However, they are relatively small (e.g.,
$\sim 1\%$ at $k\gsim 0.2\kMpch$). There are no measurable difference
at very long waves with $k\lsim 0.05\kMpch$. \footnote{The limited
  force resolution $\epsilon =1\Mpch$ for the $4\Gpch$ box and for
  smaller boxes used for Figure~\ref{fig:powerRSD} affect
  (underestimate) the redshift distortions at large wavenumbers
  $k\gsim 0.2\kMpch$. For a much better resolution of
  $\epsilon =0.25\Mpch$ and box size $L=1\Gpch$ we find that
  parameters are slightly different: $A=1.405$,
  $k_0=1.84h{\rm Mpc}^{-1}$, and $k_1=0.55h{\rm Mpc}^{-1}$. This approximation gives errors less than 0.5\% for $k<0.35\kMpch$.}

\section{Covariance matrix of the power spectrum}
\label{sec:Cov}
The covariance matrix $C(k,k^\prime)$ of the power spectrum given in
eq.(\ref{eq:cov}) is one of the main statistics required for
detailed analysis of observational survey data and estimates of
cosmological parameters \citep[see
e.g.,][]{Anderson2012,Sanchez2012,Dodelson2013,Percival2014}. It is
very difficult to estimate the covariance matrix using simulations
because thousands of realisations are required in order to produce
accurate measurements
\citep[e.g.,][]{Taylor2013,Percival2014,GLAM}. This is also a quantity
that strongly depends on the computational box-size. So, it is
important to understand how to handle $C(k,k^\prime)$ obtained from
finite-volume simulations
\citep[e.g.,][]{Gnedin2011,Li2014,Mohammed2014,Bertolini2016}.

The diagonal and off-diagonal components of the covariance matrix have
different nature and different magnitudes. The diagonal components
$C(k,k)$ are defined mostly by the Gaussian noise associated with the
finite number of Fourier harmonics found in each bin used to estimate the
power spectrum. As such, we can write:
\begin{equation}
 C^G(k,k) = \alpha\frac{2}{N_h}P^2(k),\quad N_h = \frac{4\pi k^2\Delta k}{(2\pi/L)^3},
\label{eq:gauss}
\end{equation}
where $N_h$ is the number of harmonics in a $[k,k+\Delta k]$ bin and
the coefficient $\alpha$ takes into account the filtering due to the
binning process. For the Near Grid Point (NGP) binning $\alpha =1$,
and $2/3$ for the CIC binning.  Note that the magnitude of the
diagonal components is proportional to the volume of the simulations,
i. e.,
\begin{equation}
 Cov(k,k) \propto L^{-3}.
\label{eq:covscale}
\end{equation} 
Nonlinear clustering affects the diagonal components at large
wavenumbers $(k\gsim 0.2\kMpch)$ making them larger than the simple
shot-noise estimates. However, the nonlinear terms also scale with
volume \citep{GLAM}.

\makeatletter{}\begin{figure}
\centering
\includegraphics[clip, trim= 0.5cm 5cm 0.5cm 2.5cm, width=0.49\textwidth]
{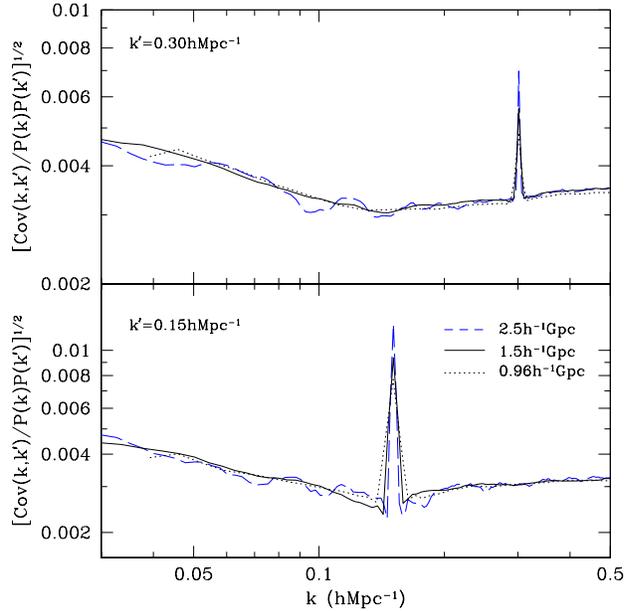}
\caption{Two slices of the dark matter covariance matrices
  $Cov(k,k^{\prime})$ in simulations with different box sizes as
  indicate the labels in the bottom panel. All covariance matrices were rescaled
  to $1.5h^{-1}Gpc$  box-size by multiplying $Cov(k,k^{\prime})$ by the
  ratio of volumes. Without this re-scaling the difference between the
  covariance matrices is very large.}
\label{fig:Covar}
\end{figure}
The non-diagonal components $C(k,k^\prime)$ have much smaller
amplitudes but there are many more of them as compared with the
diagonal ones. So, the off-diagonal componentes are still
important. Detailed analysis of these components was presented in
\citet{GLAM} who found that they also scale with the computational
volume. Here we present a couple of examples of the behaviour of the
covariance matrix in the domain of the BAO peaks.

Figure~\ref{fig:Covar} presents two slices of the dark matter
covariance matrices $Cov(k,k^{\prime})$ in simulations with different
box sizes. All covariance matrices were rescaled to the
$1.5h^{-1}Gpc$ box-size by multiplying $Cov(k,k^{\prime})$ by the ratio of
volumes. The covariance matrix of the $2.5\Gpch$ simulations (A2.5c)
was additionally scaled up by 10\%. Without this re-scalings the
difference between the covariance matrices is very large: factor
$(2.5/0.96)^3\approx 18$ between simulations with $2.5\Gpch$ and
$1\Gpch$. The large level of noise of the covariance matrix for the $2.5\Gpch$
simulations is due to the fact that the level of the signal is very low
due to the large box-size.

These results make the rescaling of the covariance matrices easy: one can rescale the
covariance matrix proportionally to the ratio of the volumes. In the next section we will 
discuss in detail the covariance corrections due to super sample modes.

\section{Super Sample Covariance}
\label{sec:SSC}
In this paper we have addressed so far the impact of missing
long-waves by studying the scaling of different quantities such as
correlation function, PDF, halo abundance, power spectrum, and
covariance matrix with the box-size. As the box-size increases, more
and more long-waves are incorporated into the simulations. The
extrapolation of these quantities to the limit of infinitely large
boxes gives us the estimates of those true quantities. These
convergence studies are the traditional way of treating situations
like that. This would work if the observational sample is very large
and we measure a fair volume fraction of the Universe. Indeed, the
current and future galaxy surveys have very large volumes. For
example, the effective volume of DESI or Euclid will be
$\sim 50\Gpch^3$,which roughly corresponds to the volume of a
simulation box with $L\sim 3.7\Gpch$.

There is another approach to the problem of waves longer than the
observed sample or computational volume that aims to estimate their
effects using a simplified model.  The key assumption of this approach
is that waves that are longer that $L$ can be considered to have a
constant (background) density $\delta_b$ inside the simulation box
\citep[see e.g.][]{Baldauf2011,TakadaHu,SuperScale,Baldauf2016}. These
very long-waves affect the growth of fluctuations inside a given box
$L$ that is extracted from a density field that does not have any
missing long-waves. There is an obvious question regarding the
accuracy of the approximation: as we saw in Section~\ref{sec:missing}
most of the missing power is in waves that are only twice longer than
the box-size $L$, and, thus cannot be considered to be constant inside
the computational box.  Let's ignore for now this question and see
what the {\it Super Scale Covariance} (SSC) approach predicts for the
covariance matrix.

The covariance matrix can be written as a sum of three terms: the
Gaussian contribution given by eq.(\ref{eq:gauss}), the nonlinear term
related with the tri-spectrum of perturbations, and the contribution of
waves longer than $L$:
\begin{equation}
 Cov_{ij}  = Cov^G(k_i,k_j)\delta_{ij} + Cov^T(k_i,k_j) + Cov^{SSC}(k_i,k_j).
\end{equation}
The first two terms scale with the volume of the simulation
$Cov^{G,T}\propto L^{-3}$ as discussed in Section~\ref{sec:Cov}. These
two terms are estimated from finite-box simulations. For this reason
we combine these two terms together and refer to the sum as
$Cov^{\rm Box}_{ij}$.

The SSC term is due to the response of the power spectrum $P(k)$ to the background density
change $\delta_b$ in the box $L$, i. e.,
$\delta P(k) = (dP(k)/d\delta_b)\delta_b$. Averaging over the
distribution of $\delta_b$ gives an estimate of the SSC covariance term \citep{TakadaHu,Li2014,Wagner2015}:
\begin{equation}
  Cov^{SSC}_{ij} \approx \sigma^2_L\frac{\partial\ln P_i}{\partial\delta_b}\frac{\partial\ln P_j}{\partial\delta_b}P_iP_j,
\end{equation}
where $\sigma_L$ is the $rms$ of $\delta_b$ as measured in boxes of
size $L$.  On large scales (small $k$) the response function
$d\ln P(k)/d\delta_b$ changes relatively slowly
\citep[e.g.,][]{TakadaHu,SuperScale,Mohammed2017}, and its magnitude
depends on how the power spectrum is measured. If $P(k)$ is measured
with respect to the local density of the simulation box $1+\delta_b$
then the effect is substantially smaller as compared with that when
the clustering is relative to the true background density. For large
galaxy surveys the clustering is relative to the average density of
the galaxy sample in the observed volume. To mimic the observations we
can always mimic the same in our simulations.

In the limit of small $k$ the response function can be written as \citep[e.g.,][]{Mohammed2017}:
\begin{equation}
  \frac{\partial\ln P}{\partial\delta_b} = \frac{5}{21}-\frac{1}{3}\frac{d\ln P}{d\ln k}\approx 0.57;
  \label{eq:respf}
\end{equation}
where this estimate is given for the power spectrum with slope -1,
which is the typical value for the long-waves $k=(0.1-0.3)\kMpch$. Note that if the
overall density is used for the background, then the first factor $5/21$  in eq.~\ref{eq:respf}
should be replaced with $41/21$ and the response function will value $\approx 2.3$.

We  estimate the $rms$ of $\delta_b$ fluctuations using a series of
$4\Gpch$ GLAM simulations with $1500^3$ particles and $3000^3$ mesh. Each
simulation was split in either $500\Mpch$ or $1\Gpch$ sub-boxes, and the
total density of each sub-box was used to find $\sigma_L$. As expected, the results
are accurately fitted by a power-law with the slope -2:
\begin{equation}
\sigma_L = \frac{1.43\times 10^{-3}}{L_{\rm Gpc}^2}, \quad L_{\rm Gpc}\equiv\frac{L}{1\Gpch}
\end{equation}
Now we estimate the impact of SSC covariance term on the normalized
covariance matrix, i. e.,
\begin{eqnarray}
  \frac{C_{ij}}{P_iP_j} &=&\frac{C^{\rm Box}_{ij}}{P_iP_j}
 +\sigma^2_L\frac{\partial\ln P_i}{\partial\delta_b}\frac{\partial\ln P_j}{\partial \delta_b}\\
                        &\approx& \frac{C^{\rm Gpc}_{ij}}{P_iP_j}
\frac{1}{L_{\rm Gpc}^{3}}+\left[\frac{0.0285}{L_{\rm Gpc}}\right]^4,
\label{eq:SSCcorr}
\end{eqnarray}
where $C^{\rm Gpc}_{ij}$ is the box covariance matrix measured for
$L=1\Gpch$ and $L_{\rm Gpc}$ is the box size in units $\Gpch$.
 
This relation can be used to estimate the correction to the covariance
matrix in simulations with a given box-size due to the super-sample
modes. For example, the covariance matrix estimated for $L=1.5\Gpch$
in Figure~\ref{fig:Covar} is
$[Cov_{ij}/P_iP_j]^{1/2}\approx (3-4)\times 10^{-3}$ for non-diagonal
components in a wide range of wavenumbers
$0.05\kMpch < k < 0.5\kMpch$. Thus, for these simulations the 
covariance matrix corrected by the SSC terms is given by
\begin{equation}
 \left[\frac{Cov_{ij}^{\rm correct}}{P_iP_j}\right]^{1/2} = 
      \left[\frac{Cov_{ij}^{\rm Box}}{P_iP_j}\right]^{1/2}\left[1+(4-7)\times 10^{-3}\right].
\end{equation}
The correction is about 0.5\%, which is small, but can be relevant
for some very sensitive applications. The estimate for
$4\Gpch$ simulations shows a 0.2\% correction.

The situation becomes totally different if the true background density is used
for the power spectrum estimates of a small computation box. For example, a box with
$L=500\Mpch$ is often used in the literature for SSC estimates
\citep[e.g.,][]{Li2014,Wagner2015,Mohammed2017}. In this case the covariance matrix
is dominated by the super-sample modes. Indeed, for this case our
estimates show that $[Cov_{ij}/P_iP_j]^{1/2}$ almost doubles due to
the SSC corrections. 

So, SSC corrections can be quite important for
surveys with small volumes or for simulations with small box
size. However, they are small for studies of galaxy clustering
statistics with effective volumes of tens of Gpc$^3$.

\makeatletter{}\section{Conclusions and Summary}
\label{sec:concl}
Modern galaxy surveys encompass larger volume of space prompting the
theory to make a careful analysis of the effects of very long waves on the
observable statistics such as the power spectrum, the correlation
functions, PDF, covariances, and the abundances of rich clusters of galaxies.
Cosmological simulations play a key role in this analysis.  It is
routinely assumed that a single cosmological simulation box must cover the
volume of the whole observable catalog. This significantly complicates the
theoretical analysis and makes nearly impossible to perform thousands
of realizations of mock galaxy samples required for estimates of
systematics and errors. We challenge this trend and make extensive
analysis of the effects due to the finite box-size of the cosmological simulations.

We argue that for most of the types of analysis of large-scale surveys a computational
volume of $L\sim (1-1.5)\Gpch$ is sufficient. In order to produce
mock observed catalogs, these finite-volume simulations should be
periodically replicated to fill the required observational
volume. We also show that no corrections are required to the average power spectrum and
correlation function, PDF and halo abundances, due to the effect of missing long-waves in a
simulation box. On the other hand, the covariance matrices should be
scaled down proportionally to the volume of the observations and, if
necessary, corrected for the super-sample modes as given in eq.(\ref{eq:covscale})
and eq.(\ref{eq:SSCcorr}).

\medskip
Here is the summary of our main results:

\medskip\noindent -- Defects of box replications can be readily
remedied by a combination of sufficiently large $\sim 1\Gpch$
simulations and rotation of the boxes before building mock galaxy
catalogs,

\medskip\noindent -- The missing power of finite-box simulations (due
to waves longer than the computational box) dramatically declines with
increasing box-size $L$, and becomes extremely small
$\sigma\lsim 0.01$ for $L\gsim 1\Gpch$. Most of the missing power is
in waves that are slightly longer than the box-size: about 90\% of the
missing power is in waves with wavelengths $(1-1.5)L$,

\medskip\noindent -- Corrections to the abundance of halos and galaxies are
extremely small and can be neglected for computational volumes larger than $1\Gpch$,

\medskip\noindent -- The average power spectra of dark matter
fluctuations show remarkable lack of dependance on the size of the
computational box. We clearly detect some decline of the amplitude of
fluctuations for small $(250-500)\Mpch$ boxes, but it is small:
$(1-1.5)$\% for the smallest $250\Mpch$ simulation that we
studied. There are no visible effects for simulations with
$L>1\Gpch$ with upper limits of $\sim 0.5$\% for extremely long-waves
with $k=(0.008-0.05)\kMpch$ and less than $\sim 0.1$\% for waves in
the BAO domain with $k=(0.07-0.3)\kMpch$.

\medskip\noindent -- The covariance matrix of the dark matter power
spectra scales proportionally to the computational volume. This well
known result \citep[e.g.,][]{TakadaHu,Wagner2015,GLAM} is important
for using mock galaxy catalogs: the covariance matrix must be scaled
down to match the observational sample. The SSC correction to the
covariance matrix is expected to be $\sim 0.5$\% for observational
samples with effective volume $3(\Gpch)^3$ (box-size $L=1.5\Gpch$),
and becomes negligible when the observational sample increases to
$\sim 50(\Gpch)^3$ expected for DESI/Euclid and LSST surveys.

\medskip\noindent -- The most stringent constraints on the simulation
volume are coming from the requirement that mock catalogs should
reproduce not only the correct power spectra, but also the correlation
functions \citep{Sirko2005,Klypin2013}. We find that the correlation
functions for $L\lsim 500\Mpch$ are qualitatively incorrect. For
example, for $L= 500\Mpch$ the dark matter correlation function is
zero at $R\approx 85\Mpch$ where it must be positive. For
$L= 300\Mpch$ the correlation function is negative for the whole
domain of the BAO peak ($R\approx 100\Mpch$).  However, the effect
quickly becomes very small with increasing volume and is negligible
for $L\gsim 1\Gpch$.

Based on the work presented in this paper we conclude that a
simulation box of $L\sim (1-1.5)\Gpch$ is large enough to fulfil most
of the science requirements, in the fields of large-scale structure,
weak-lensing and cosmological parameters, of the upcoming new
generation of large redshift surveys.

\section*{Acknowledgements}
We thank J. Peacock and M. Schmittfull for discussions and comments.
A.K. acknowledges support of the Fulbright Foundation,  support of
the Instituto de Astrofisica de Canarias, La Laguna, and the Severo Ochoa scholarship. A.K.  and
F.P. acknowledges support from the Spanish MINECO grant
AYA2014-60641-C2-1-P.  F.P. wants to thank the support and hospitality
of the ICC at Durham University where part of this work was
completed. The new GLAM simulations presented in this paper were done
at the Barcelona Supercomputer Center (Spain) and the DiRAC Data
Centric system at Durham University, operated by ICC on behalf of the
STFC DiRAC HPC Facility. We thank New Mexico State University (USA)
and Instituto de Astrofisica de Andalucia CSIC (Spain) for hosting the
skiesanduniverse.org site for cosmological simulation products.

\bibliography{Box}
\bibliographystyle{mn2e}
\end{document}